\documentclass[twoside,showpacs,superscriptaddress,twocolumn,floatfix,a4paper,prl]{revtex4}

\usepackage{color}
\usepackage{graphicx}
\usepackage[utf8x]{inputenc}
\usepackage[T1]{fontenc}

\begin{document}

\title{Linear-optical programmable quantum router}

\author{Karel Lemr}
\affiliation{RCPTM, Joint Laboratory of Optics of Palacký University and Institute of Physics of Academy of Sciences of the Czech Republic, 17. listopadu 12, 772 07 Olomouc, Czech Republic}
\author{Antonín Černoch}
\affiliation{Institute of Physics of Academy of Sciences of the Czech Republic, Joint Laboratory of Optics of PU and IP AS CR, 
   17. listopadu 50A, 779 07 Olomouc, Czech Republic}

\date{\today}

\begin{abstract}
This paper presents a scheme for linear-optical implementation of a programmable quantum router. Polarization encoded photon qubit is coherently routed to various spatial modes according to the state of several control qubits. In our implementation, the polarization state of the signal photon does not change under the routing operation. We also discuss generalization of the scheme that would allow to obtain signal dependent routing.
\end{abstract}


\maketitle
%\section{Introduction}
Quantum information processing is a combined field of physics and information science \cite{nielsen}. It is dedicated to improve capability to transmit and process information by employing the laws of quantum physics. In many aspects, the quantum information processing is inspired by its classical analogue. Such analogy is observable for instance on distribution of quantum information, where one would use networks very similar to classical information networks. This similarity manifests itself also in various components needed to build both classical and quantum information networks \cite{barenco}.

An important building block of classical information networks are the so-called routers -- devices used to direct the information from the source to its intended destination \cite{medhi_routing}. The more complex the network is, the more pronounced is the need for correct routing of the information. Classical routers are indeed densely used and they are subject to intensive applied research \cite{duguay69router,jackel90router,ruiqiang11router}. Their quantum analogues -- the quantum routers -- are however still in the phase of basic research.

Quantum routing can be achieved for instance by employing matter-light interaction \cite{zueco09router,aoki09router,hoi12router}. These techniques might however prove experimentally demanding. An all-optical quantum router has also been reported on in Ref.~\cite{matthew12switch}, but this device use only classical information as a control for routing. Very recently, X.-Y.~Chang {\it et al.} proposed a strategy for linear-optical quantum routing based on the entanglement between signal and control qubit \cite{chang12router}. In their proposal the routing control is represented by a qubit. However, the information stored in the signal qubit collapses depending on the measurement performed on the control qubit. Also the requirement to maintain entanglement between the routed and control qubit may be limiting.

\begin{figure}
\includegraphics[scale=1]{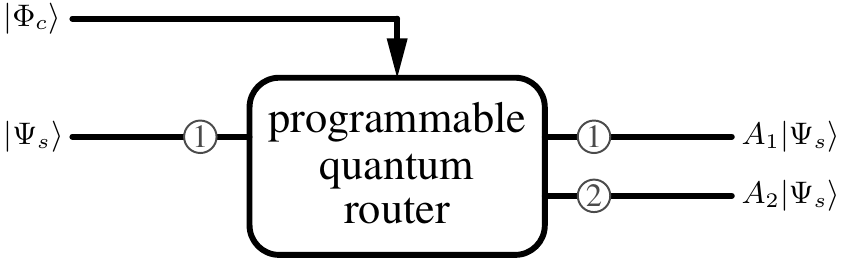}
\caption{Conceptual scheme of a programmable quantum router. The input signal qubit $|\Psi_s\rangle$ initially in spatial mode 1 is coherently routed to output modes 1 and 2 depending on the state of the control qubit(s). Since the quantum information itself is stored in other then spatial degree of freedom (we use polarization encoding), it is not modified by the routing procedure.}
\label{fig:concept}
\end{figure}

In this paper, we propose a scheme for implementation of a linear-optical programmable quantum router (see conceptual scheme in Fig. \ref{fig:concept}). As the title suggests, the signal qubit is routed according to the state of control qubit(s). Since the device is ``quantum'', it can route the signal qubit into a coherent superposition of output modes. Our implementation combines the benefits of the above mentioned implementations. It is all-optical, fully quantum (routing control is achieved by means of a qubit) and it does not require any previous entanglement between signal and control qubit. On top of that, the information stored in the signal qubit does not change under the routing procedure. Since the devices is based solely on linear-optical components, it can be implemented experimentally with presently available technologies \cite{kok_linear}.

We employ two distinct types of quantum information encoding. Polarization encoding is used for implementing the signal qubit state and spatial mode encoding is then used for routing. The polarization state is maintained independently on the actual routing.
%
%\section{Principle of operation}
%
\begin{figure}
\includegraphics[scale=1]{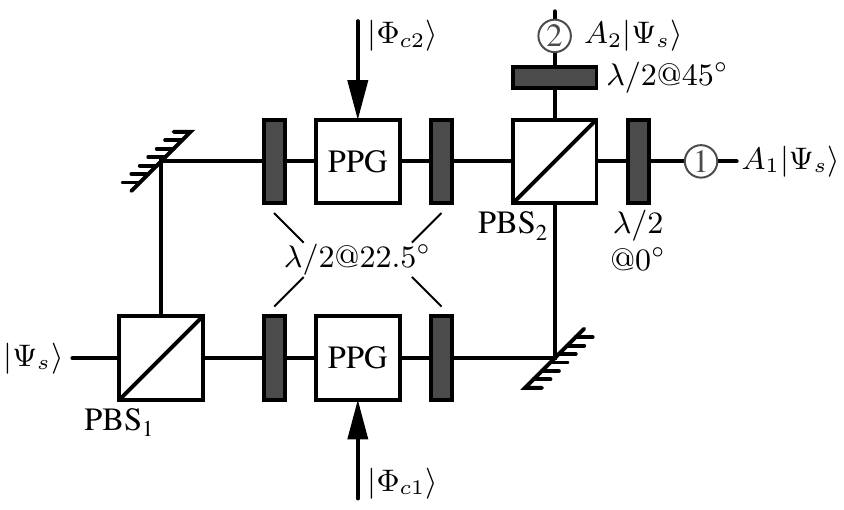}
\caption{Scheme of the linear-optical implementation of a programmable quantum router. PPG denotes the programmable phase gate, $\lambda/2$ denotes half-wave plate and PBS stands for polarizing beam splitter (horizontal linear polarization is transmitted while vertical linear polarization is reflected). The signal input qubit is encoded into polarization state of the signal photon. After interacting with control qubits in the PPG, the signal photon with polarization state is unchanged is routed to a coherent superposition of the two output ports labelled 1 and 2.}
\label{fig:scheme}
\end{figure}
The hereby proposed linear-optical implementation of the quantum router is depicted in Fig. \ref{fig:scheme}. As already stated above, the signal qubit is stored in polarization state of the signal photon
\begin{equation}
|\Psi_s\rangle_1 = \alpha|H\rangle + \beta|V\rangle,
\end{equation}
where $|H\rangle$ denotes the horizontal and $|V\rangle$ vertical linear polarization states and $|\alpha|^2 + |\beta|^2 = 1$. Index 1 expresses the fact that initially the signal qubit is in spatial mode 1.
Similarly the routing control is stored in polarization state of control photons $|\Phi_{c1}\rangle$ and $|\Phi_{c2}\rangle$
\begin{equation}
\label{eq:c1_c2}
|\Phi_{c1}\rangle = |\Phi_{c2}\rangle = \frac{1}{\sqrt{2}}\left(|H\rangle + \mathrm{e}^{i\phi_1}|V\rangle\right).
\end{equation}
Note that both the control qubits are in the same state.

The router works as follows: the signal is brought to the input port of the first polarizing beam splitter PBS$_1$. There the horizontal and vertical polarization components split and proceed by separate arms (horizontal polarization is transmitted while vertical is reflected). In these arms, both the polarization components undergo a programmable phase gate (PPG) \cite{vidal02gate} enveloped by Hadamard gates placed in front and behind the PPG. The Hadamard gates are easily implemented on polarization encoded qubits just by using a half-wave plate rotated by $22.5$ deg. with respect to horizontal polarization direction. In the PPG, the signal interacts with the control qubits $|\Phi_{c1}\rangle$ and $|\Phi_{c2}\rangle$.

These gates perform the following unitary transformation on the signal state
\begin{equation}
U_\mathrm{PPG}(\phi_1) |H\rangle = |H\rangle,\quad U_\mathrm{PPG}(\phi_1) |V\rangle = \mathrm{e}^{i\phi_1}|V\rangle,
\end{equation}
where $\phi_1$ is the phase shift in the control qubit as defined in Eq. (\ref{eq:c1_c2}). The phase shift $\phi_1$ of the control qubit is imposed to the vertical component of the signal qubit. Linear-optical PPG works probabilistically with maximum success probability of $\frac{1}{2}$. Its successful operation is conditioned on detection of only one photon in the control mode (further details can be found in \cite{micuda08gate,mikova12gate}).

One can easily verify, that the above mentioned procedure renders the signal state to the form 
\begin{eqnarray}
|\Psi_s\rangle_1 & \rightarrow & \frac{\left(1+\mathrm{e}^{i\phi_1}\right)}{2}\left({\alpha}|H\rangle_1 + {\beta}|V\rangle_2\right) \\\nonumber
& + & \frac{\left(1-\mathrm{e}^{i\phi_1}\right)}{2}\left({\alpha}|V\rangle_1 + {\beta}|H\rangle_2\right),
\end{eqnarray}

After that the signal is recombined on the second polarizing beam splitter PBS$_2$. The entire signal transformation now reads
\begin{eqnarray}
|\Psi_s\rangle_1 & \rightarrow & \frac{1}{2}\left[\left(1+\mathrm{e}^{i\phi_1}\right)\left(\alpha|H\rangle_1 - \beta|V\rangle_1\right)\right. \\\nonumber
& + & \left.\left(1-\mathrm{e}^{i\phi_1}\right)\left(\beta|H\rangle_2 + \alpha|V\rangle_2\right)\right].
\end{eqnarray}

Simple corrections are needed in both output modes to revert the signal state back to its original form. Namely in the first output mode, one needs to perform a $\pi$ phase shift between horizontal and vertical polarizations. This is achieved by a half-wave plate rotated by $0$ deg. In the second output mode, the undesired polarization $H\leftrightarrow V$ swap is compensated by another half-wave plate with optical axis rotated by $45$ deg. After these correction are performed, the entire routing operation can be expressed in the form
\begin{equation}
\label{eq:routing}
|\Psi_s\rangle_1 \rightarrow A_1|\Psi_{s}\rangle_1 + A_2|\Psi_{s}\rangle_2,
\end{equation}
where
\begin{equation}
A_1 = \frac{1}{2}\left(1+\mathrm{e}^{i\phi_1}\right),\quad A_2 = \frac{1}{2}\left(1-\mathrm{e}^{i\phi_1}\right).
\end{equation}
The routing action is clearly observable from Eq. (\ref{eq:routing}). There is a clear formal resemblance between the routing transformation and the action of a polarization independent beam splitter. One can explore this similarity by introducing transmissivity $T$ and reflectivity $R$ and describing the router in terms a programmable beam splitter
\begin{equation}
\label{eq:routing_bs}
|\Psi_s\rangle_1 \rightarrow \sqrt{T}|\Psi_{s}\rangle_1 + \sqrt{R}\mathrm{e}^{-i\frac{\pi}{2}}|\Psi_{s}\rangle_2,
\end{equation}
where
\begin{equation}
T = \frac{1}{2}\left(1+\cos\phi_1\right),\quad R = 1 - T = \frac{1}{2}\left(1-\cos\phi_1\right).
\end{equation}
Note that by means of the control qubits $|\Phi_{c1}\rangle$ and $|\Phi_{c2}\rangle$ one can fully tune the intensity transmissivity $T$ from 0 to 1. On the other hand, there is no control over the phase shift between two output modes which is fixed to $-\frac{\pi}{2}$. This limitation does not burden applications where the router is used for simple signal directing. Since two PPGs are employed, the overall success probability of the router is $\frac{1}{4}$.

\begin{figure}
\includegraphics[scale=1]{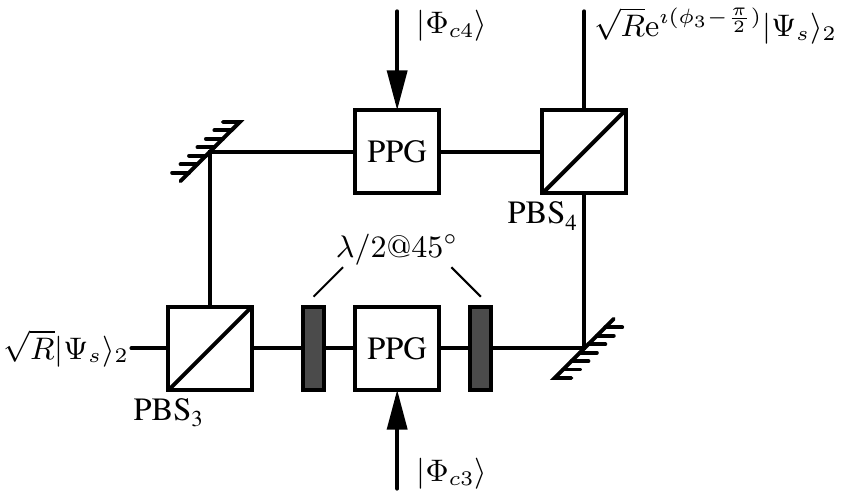}
\caption{Additional block of two PPGs used to set required phase shift between the two output modes of the quantum router. Components are labelled as in Fig. \ref{fig:scheme}}
\label{fig:addblock}
\end{figure}
If there is also a need for programmable control of the phase shift additional apparatus has to be added to one of the output modes. Suppose it is added to the second output mode. This additional block of optical components (depicted in Fig. \ref{fig:addblock}) is composed again of two polarizing beam splitters (PBS$_3$ and PBS$_4$) forming a interferometer similar to the one formed by PBS$_1$ and PBS$_2$. There are two PPGs, one in each arm of this interferometer, but this time, there are no Hadamard gates present. Instead of that, the arm that processes horizontal component of the signal state is equipped by two half-wave plates rotated by $45$ deg. These wave plates encompass the PPG and swap the horizontal polarization to vertical and then back again. Direct calculation reveals that the action of this entire additional apparatus shifts the signal state in the second arm by the phase $\phi_3$ corresponding to the phase in the definition of control qubit states $|\Phi_{c3}\rangle$ and $|\Phi_{c4}\rangle$
\begin{equation}
|\Phi_{c3}\rangle = |\Phi_{c4}\rangle = \frac{1}{\sqrt{2}}\left(|H\rangle + \mathrm{e}^{i\phi_3}|V\rangle\right).
\end{equation}
Note that the control qubits $|\Phi_{c3}\rangle$ and $|\Phi_{c4}\rangle$ share the same state.

Including the additional block, the overall action of the router represent the transformation
\begin{equation}
|\Psi_s\rangle_1 \rightarrow \sqrt{T}|\Psi_{s}\rangle_1 + \sqrt{R}\mathrm{e}^{i\left(\phi_3-\frac{\pi}{2}\right)}|\Psi_{s}\rangle_2.
\end{equation}
This way the router can control not only the intensity splitting of the signal, but also the phase shift between the output modes. The price to pay for this additional degree of freedom is decreased success probability which is now $\frac{1}{16}$.

%\section{Generalized action}
The action of the router can be generalized to provide more signal dependent complex routing. For this task, the conditions imposed on the control qubit states
\begin{equation}
|\Phi_{c1}\rangle = |\Phi_{c2}\rangle,\quad|\Phi_{c3}\rangle = |\Phi_{c4}\rangle
\end{equation}
is abandoned and independent control qubits in the form
\begin{equation}
|\Phi_{cN}\rangle = \frac{1}{\sqrt{2}}\left(|H\rangle+\mathrm{e}^{i\phi_N}|V\rangle\right),\quad N = 1,2,3,4
\end{equation}
are used. The router then performs the following generalized transformation
\begin{eqnarray}
|H\rangle_1 & \rightarrow & \mathrm{e}^{\frac{i\phi_1}{2}}\left(\sqrt{T_1} |H\rangle_1 + \sqrt{R_1}\mathrm{e}^{i\left(\phi_3-\frac{\pi}{2}\right)} |H\rangle_2\right) \\\nonumber
|V\rangle_1 & \rightarrow & \mathrm{e}^{\frac{i\phi_2}{2}}\left(\sqrt{T_2} |V\rangle_1 + \sqrt{R_2}\mathrm{e}^{i\left(\phi_4-\frac{\pi}{2}\right)} |V\rangle_2\right),
\end{eqnarray}
where indexes behind brackets denote spatial modes as usual and $T_N$ and $R_N$ for $N=1,2$ depend on the control states $|\Phi_{c1}\rangle$ and $|\Phi_{c2}\rangle$
\begin{equation}
T_N = \frac{1}{2}\left(1+\cos\phi_N\right),\quad R_N = 1 - T_N =  \frac{1}{2}\left(1-\cos\phi_N\right).
\end{equation}

%\section{Conclusions}
This paper brings forward a proposal for experimentally feasible programmable quantum router based on linear optics. The device works probabilistically with success probability of $1/4$ resp. $1/16$ if control over phase between the output modes is required. It is capable of routing a polarization signal qubit to two distinct spatial output modes depending on the state of the control qubits. Even though it requires 2--4 control qubits, it only uses two-qubit PPG which has already been achieved experimentally. Finally a generalisation of the scheme is discussed showing how to modify the router to perform signal dependent routing.

%\section{Acknowledgment}
The authors also gratefully acknowledge the support by the Operational Program Research and Development for Innovations -- European Regional Development Fund (project CZ.1.05/2.1.00/03.0058  and the  Operational Program Education for Competitiveness -- European Social Fund (project CZ.1.07/2.3.00/20.0017 of the Ministry of Education, Youth and Sports of the Czech Republic. This paper is dedicated to our wives and families.

\end{document}